\documentclass[prl,twocolumn,showpacs]{revtex4}

\usepackage{graphicx}
\usepackage{amsmath}

\begin{document}

\title{Exciting Collective Oscillations in a Trapped 1D Gas}

\author{Henning Moritz, Thilo St\"{o}ferle, Michael K\"{o}hl$^\dag$ and Tilman Esslinger}

\affiliation{Institute of Quantum Electronics, ETH Z\"{u}rich
H\"onggerberg, CH--8093 Z\"urich, Switzerland}

\date{\today}

\begin{abstract}
We report on the realization of a trapped one dimensional Bose gas
and its characterization by means of measuring its lowest lying
collective excitations. The quantum degenerate Bose gas is
prepared in a 2D optical lattice and we find the ratio of the
frequencies of the lowest compressional (breathing) mode and the
dipole mode to be $(\omega_B/\omega_D)^2\simeq3.1$, in accordance
with the Lieb-Liniger and mean-field theory. For a thermal gas we
measure $(\omega_B/\omega_D)^2\simeq4$. By heating the quantum
degenerate gas we have studied the transition between the two
regimes. For the lowest number of particles attainable in the
experiment the kinetic energy of the system is similar to the
interaction energy and we enter the strongly interacting regime.
\end{abstract}


\maketitle

An ultra cold Bose gas in one spatial dimension is different from
its two or three dimensional counterparts. One striking example is
that Bose-Einstein condensation does not occur at finite
temperature in a homogeneous one dimensional system. In an
interacting Bose gas the constraint to one dimension leads to
another remarkable and counterintuitive property. With decreasing
atomic density the interactions become increasingly dominant and
the character of the system changes. Assuming delta-functional
interactions, exact solutions have been found for the ground state
and the excitation spectrum of a homogeneous one dimensional Bose
gas \cite{Girardeau1960,Lieb1963}. Sparked by the prospect that
this unique model in many-body quantum physics could become
experimentally accessible there has recently been a wave of
theoretical interest in trapped 1D gases. Assuming elongated
trapping geometries in which the radial atomic motion is confined
to zero point oscillation different physical regimes could be
identified
\cite{Olshanii1998,Ho1999,Petrov2000,Girardeau2001,Menotti2002,Lieb2003,Pedri2003}.
The stringent requirements for testing these models have so far
not been reached in experiments.

A trapped 1D gas is characterized by a single parameter $\gamma$
which is the ratio between the interaction energy and the kinetic
energy of the ground state: $\gamma=\frac{m g_{1D}} {\hbar^2
n_{1D}}$, with $m$ being the atomic mass, $g_{1D}$ the 1D coupling
constant and $n_{1D}$ the 1D density. For high densities ($\gamma
\ll 1$) the system is weakly interacting, a regime which can be
described by mean-field theory, and in harmonically confined 1D
systems Bose-Einstein condensation is possible. When the 1D
density is lowered the kinetic energy of the ground state is
reduced and may get smaller than the interaction energy thereby
transforming the gas into a strongly interacting system, where the
longitudinal motion of the particles is highly correlated. For
$\gamma \gtrsim 1$ the correlation length $l_c=\hbar/\sqrt{m
n_{1D} g_{1D}}$ becomes shorter than the mean interparticle
distance. In the limit $\gamma \rightarrow \infty$, often referred
to as the Tonks-Girardeau regime, a Bose gas acquires fermionic
properties due to the strong repulsive interactions. The different
regimes associated with the parameter $\gamma$ are characterized
by the excitation spectrum which can be probed by measuring the
frequencies of collective oscillations
\cite{Ho1999,Menotti2002,Pedri2003}. In the two limiting cases of
weak and strong interactions the ground state properties and the
excitation spectrum are explicitly known, whereas in the crossover
regime many unknown issues remain to be resolved for the confined
system.

\begin{figure}[htbp]
  \includegraphics[width=.6\columnwidth,clip=true]{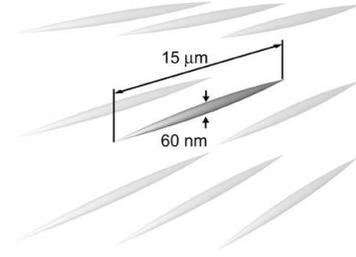}
  \caption{The geometry and size of trapped 1D gases in a two dimensional optical lattice. The spacing between the 1D tubes
  in the horizontal and vertical direction is 413\,nm.}
  \label{fig1}
\end{figure}

The 1D regime is reached, when the condition
\begin{eqnarray}
\mu, k_B T \ll \hbar \omega_r
\end{eqnarray}
is fulfilled, where $\omega_r$ denotes the radial trapping
frequency and thus the strength of the radial harmonic
confinement, $\mu$ the chemical potential, and $T$ the
temperature. If the ground state extension in the radial direction
$a_r=\sqrt{\hbar/m \omega_r}$ is much larger than the
characteristic radius of the interatomic potential, the 1D
coupling strength $g_{1D}$ can be expressed in terms of the 3D
scattering length $a$ through $g_{1D}=-\frac{2 \hbar^2}{m a_{1D}}$
with $a_{1D}\simeq -a_r^2/ a$ being the 1D scattering length
\cite{Olshanii1998}. The necessity for low densities and therefore
low atom numbers makes the experimental quest for 1D gases
challenging. A simple estimate using the Lieb-Liniger chemical
potential $\mu=2 \hbar \omega_r n_{1D} a$ shows that for a
degenerate 1D quantum gas in the mean field regime the density
must obey $n_{1D} \ll 1/a$ and the number of particles must be $N
\ll L/a \simeq 10^3-10^4$, where $L$ is the characteristic length
of the system, e.g. the Thomas-Fermi radius.

There has been significant progress towards the realization of
trapped 1D atomic gases over the past years. Both, in a
$^6$Li/$^7$Li mixture \cite{Schreck2001} and in $^{23}$Na
\cite{Gorlitz2001}, quantum degenerate gases have been created in
very elongated traps and features of one dimensional condensate
expansion were observed. Considering only the condensed fraction a
chemical potential of $\mu \gtrsim 0.5 \,\hbar \omega_r$ was
attained in these experiments. However, the thermal component,
which was a substantial portion of the gas, was in a 3D
configuration ($k_B T> \hbar \omega_r$) leaving the whole sample
in an interesting crossover regime. In a similar experimental
regime $^7$Li Bose-Einstein condensates with attractive
interparticle interactions were launched into 1D matter waveguides
forming bright matter wave solitons \cite{Khaykovich2002}.
Moreover, a Bose-Einstein condensate of rubidium has been loaded
into the ground state of a two dimensional optical lattice, where
the transverse oscillations have been frozen out but the tunneling
rate between the tubes exceeded the axial trapping frequency,
resulting in an array of strongly coupled tubes
\cite{Greiner2001b}.

In our experiment we have realized both quantum degenerate and
thermal one dimensional atomic gases with the condition (1) being
well fulfilled: for all our experiments $k_BT/\hbar \omega_r<6
\times 10^{-3}$ and $\mu/\hbar \omega_r<0.1$. The atoms are
prepared in a 2D optical lattice which offers the advantage of an
extremely tight radial confinement of only a fraction of the
optical lattice wavelength. Moreover, the geometry (see Fig.
\ref{fig1}) makes it possible to study many copies of the 1D
system at the same time, thereby circumventing problems arising
from the detection of very low particle numbers. The parameter
$\gamma$ ranges approximately from $0.4$ to $1$, which is at the
crossover from the mean field to the strongly correlated regime.

In the experiment, we collect up to $2 \times 10^9$ $^{87}$Rb
atoms in a vapor cell magneto-optical trap. After polarization
gradient cooling and optical pumping into the $|F=2, m_F=2\rangle$
hyperfine ground state the atoms are captured in a magnetic
quadrupole trap. Subsequently, we magnetically transport the
trapped atoms using a series of partially overlapping quadrupole
coils \cite{Greiner2001a} over a distance of 40\,cm into an
optical quality quartz cell, which is pumped to a pressure below
$2\times 10^{-11}$\,mbar. The transfer time is 1.5\,s and we have
observed transfer efficiencies larger than 80\% with no detectable
heating of the cloud. Finally, the linear quadrupole potential is
converted into the harmonic and elongated potential of a QUIC-trap
\cite{Esslinger1998}. Radio frequency induced evaporation of the
cloud is performed over a period of 25\,s, during which the
confinement is adiabatically relaxed to $\omega_\perp=2 \pi \times
120$\,Hz in the radial and $\omega_\|=2 \pi \times 20$\,Hz in the
axial direction. We produce almost pure Bose-Einstein condensates
of up to $3\times 10^5$ atoms. After condensation we adiabatically
change the trapping geometry to an approximately spherical
symmetry with trapping frequencies of $\widetilde{\omega}_x=2 \pi
\times 17$\,Hz, $\widetilde{\omega}_y=2 \pi \times 20$\,Hz, and
$\widetilde{\omega}_z=2 \pi \times 22$\,Hz. This reduces the peak
density by a factor 4 and allows us to load the optical lattice
more uniformly.

The optical lattice is formed by two retro-reflected laser beams,
which are derived from laser diodes at a wavelength of
$\lambda=826$\,nm. At the position of the condensate, the laser
beams overlap perpendicularly with orthogonal polarizations and
are focused to a circular waist (1/$e^2$-radius) of 105 $\mu$m and
120 $\mu$m, respectively. The frequencies of the two beams are
offset with respect to each other by 152\,MHz. The optical
potential depth $U$ is proportional to the laser intensity and is
conveniently expressed in terms of the recoil energy
$E_{rec}=\frac{\hbar^2 k^2}{2 m}$ with $k=\frac{2 \pi}{\lambda}$.
The two dimensional optical lattice produces an array of tubes,
tightly confined in the radial direction and spaced by the
periodicity of the lattice $d=\lambda/2$ (see Fig. \ref{fig1}).
The radial confinement of the tube is characterized by the radial
trapping frequency $\hbar \omega_r \simeq 2 \sqrt{s} E_{rec}$,
where $s=U/E_{rec}$ expresses the potential depth in units of
$E_{rec}$. The axial confinement of the tubes is determined by the
waists $\textrm{w}$ and intensities of the laser beams and
characterized by the trapping frequency $\hbar \omega_z \simeq 2
\sqrt{s}\frac{\lambda}{\pi \textrm{w}} E_{rec}$. We have
calibrated the potential depth of each of the optical lattice
laser beams by measuring the frequency of small amplitude dipole
oscillations along the axis of the laser beam. From the
oscillation frequency we deduce the effective mass $m^*/m$ at the
quasi-momentum $q=0$ in the band structure, which is a measure of
the potential depth of the optical lattice
\cite{Cataliotti2001,Kraemer2002}. The calibration error is
estimated to be $<10\%$.

Adiabatic loading into the ground state of the optical lattice was
achieved by ramping up the laser intensity to $U=30\,E_{rec}$ with
an exponential ramp using a time constant $\tau=75$\,ms and a
duration of $t_0=150$\,ms. We have verified experimentally that
all atoms are loaded into the lowest Bloch band of the optical
lattice by ramping down the intensity of the lattice laser beams
adiabatically and observing in the time-of-flight image, after
release from the magnetic trap, that only the lowest Brillouin
zone was occupied \cite{Greiner2001b}.

The 1D systems in the optical lattice are not perfectly isolated,
but the tubes are coupled by the tunnelling matrix element $J$.
For sufficiently deep lattice potentials, the tunnelling becomes
exponentially small and contributes only a small correction of
order $J/\mu$ to the 1D characteristics in the individual tubes.
If $J/\mu \ll 1$ locally the gas acquires 1D properties and can be
well described by a local Lieb-Liniger model, even though the
whole sample is three dimensional \cite{Pedri2003}. Experimentally
we observe the disappearance of the matter wave interference
pattern  with increasing lattice depth when the atoms are suddenly
released from the optical lattice. Higher order momentum peaks
($\pm 2 \hbar k, \pm 4 \hbar k, ...$) are usually observed at
lower laser intensities \cite{Greiner2001b}. We attribute this
loss of coherence between the individual tubes to the very small
tunnel coupling at large lattice depths, which is too small to
stabilize the global phase coherence.

The different regimes of the 1D gas in the optical lattice can be
determined from the excitation spectrum. The frequency ratio
between the lowest compressional mode (breathing mode) and the
dipole oscillation $(\omega_B/\omega_D)^2$ is a sensitive measure,
both for isolated 1D systems \cite{Menotti2002} and for atoms in
an optical lattice \cite{Pedri2003}. For a degenerate gas in the
1D mean field regime, one expects $(\omega_B/\omega_D)^2=3$,
whereas in the Tonks-Girardeau regime $(\omega_B/\omega_D)^2=4$.
The latter frequency ratio is the same for both a thermal gas and
a gas of degenerate, noninteracting fermions. In contrast, for a
three dimensional elongated condensate in the mean field regime
the ratio of the oscillation frequencies is
$(\omega_B/\omega_D)^2=5/2$
\cite{Stringari1996,Mewes1996,Chevy2002}.

\begin{figure}[htbp]
  \includegraphics[width=.75\columnwidth,clip=true]{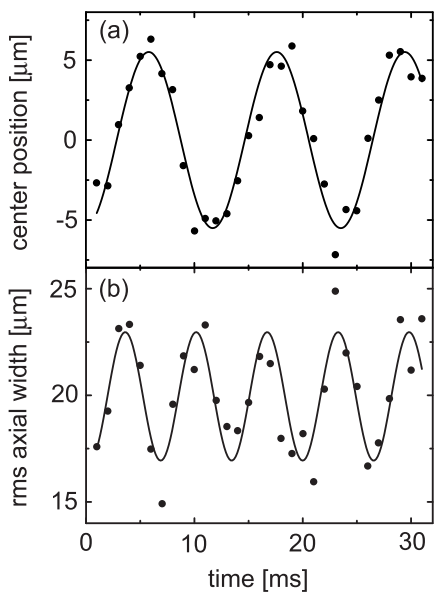}
  \caption{Dipole oscillation (a) and breathing mode (b) of a quantum degenerate one
  dimensional Bose gas. For this data set an almost pure Bose-Einstein condensate with $N=(9.8\pm 0.8)\times 10^4$ atoms was
  loaded into the optical lattice and imaged after 15\,ms of ballistic expansion.
  From the fits we obtain $\omega_D=2 \pi \times (84.6\pm 0.4)$\,Hz and
  $\omega_B=2 \pi \times (152.6 \pm 2.0)$\,Hz.}
  \label{fig2}
\end{figure}

We have measured the frequency of the collective excitations of
the atoms in the optical lattice. The breathing mode was excited
by sinusoidal intensity modulation of the optical lattice with an
amplitude of 4\,$E_{rec}$ for five cycles and a frequency of
150\,Hz, which is close to but not matching the expected frequency
of the breathing mode. At the end of the modulation period a short
(1 ms) magnetic field gradient was applied along the symmetry axis
of the 1D tubes to induce a dipole oscillation of the condensate
in the axial trapping potential. After a variable evolution time
in the combined optical and magnetic trapping potential all
confining forces were suddenly switched off \cite{remark1} and we
detected the atoms after ballistic expansion by absorption
imaging. The density distribution of the atoms was fitted by a
Gaussian to extract the position and width of the cloud. Figure 2
shows a data set of a dipole oscillation (Fig. 2a) and a breathing
mode (Fig. 2b). In order to extract the frequencies of the modes,
we have fitted an exponentially decaying sine function to the
position of the cloud and an exponentially decaying sine function
plus a linearly increasing term to the rms axial width of cloud.
The latter accounts for the observation that there is a slight
increase in the width of cloud with longer hold times, possibly
due to technical noise. The damping coefficients varied between
0\,s$^{-1}$ and 40\,s$^{-1}$ for the dipole oscillations and
between 3\,s$^{-1}$ and 60\,s$^{-1}$ for the breathing mode. We
found that the damping coefficients depend critically on the
alignment of the optical lattice.

\begin{figure}[htbp]
  \includegraphics[width=.75\columnwidth,clip=true]{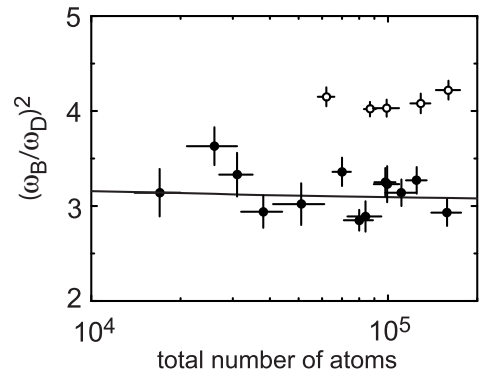}
  \caption{The measured frequency ratio $(\omega_B/\omega_D)^2$ for a Bose condensed 1D gas (solid symbols).
  The solid curve is the theoretical prediction from  \cite{Pedri2003} using the measured dipole frequency
  along the 1D tubes $\omega_z=2 \pi \times 84$\,Hz and the
  frequency $\omega$ of the slowly varying confining potential in the
  transverse direction  with $\sqrt{m/m^*} \omega=2 \pi \times 4$\,Hz.
  Averaging all measurements, independent of the atom number, we obtain a frequency ratio
  of $(\omega_B/\omega_D)^2=3.15 \pm 0.22$.  For a 1D gas
  of thermal atoms we find  $(\omega_B/\omega_D)^2=4.10 \pm 0.08$ (open circles).
  The depth of the optical lattice is $30\,E_{rec}$. The error bars reflect only the statistical
  uncertainties on the total atom number and fit errors on the frequencies.}
  \label{fig3}
\end{figure}

Figure 3 shows the measured ratio $(\omega_B/\omega_D)^2$ for a
pure 3D-Bose-Einstein condensate loaded into a 2D optical lattice
for various total atom numbers. For all data points there was no
discernible thermal cloud which allows us to estimate for the
temperature $T/T_{c,3D}<0.3$, where $T_{c,3D}$ denotes the
critical temperature for Bose-Einstein condensation in the final
magnetic trapping configuration. We compare the measured ratio
$(\omega_B/\omega_D)^2$ to the theoretical prediction of
\cite{Pedri2003} (solid line) and find good agreement over the
wide range of atom numbers investigated. For the lowest total atom
numbers $N=1.7\times 10^4$ the parameter $\gamma$ reaches unity in
the central tube of the lattice, indicating that we are in the
crossover region from the 1D mean field regime to the
Tonks-Girardeau regime. We have estimated the number of atoms in
the central tube to be 30, assuming that the overall 3D density
profile is Thomas-Fermi-like and using an effective coupling
constant $\widetilde{g}$ which is modified by the optical lattice
\cite{Kraemer2002,Kraemer2003}.

We have also loaded thermal gases into the optical lattice and
obtained an average value $(\omega_B/\omega_D)^2=4.10 \pm 0.08$
without significant dependence on total atom number $N$ and
temperature $T$ over the range of $6\times 10^4 <N<1.6 \times
10^5$ and $54\,\textrm{nK} < T < 91\,\textrm{nK}$. For thermal
clouds we have observed that the frequency of the dipole
oscillations is up to 5\% smaller than for the Bose condensed
clouds. We attribute this to the larger size of the thermal clouds
which therefore might experience anharmonic parts of the optical
potential.

\begin{figure}[htbp]
  \includegraphics[width=.75\columnwidth,clip=true]{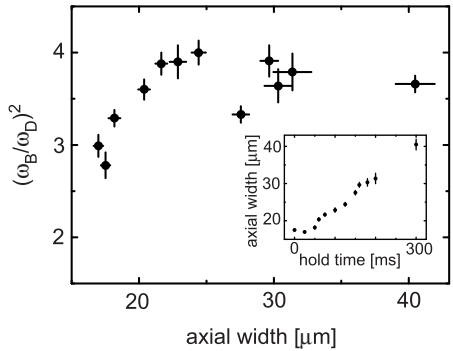}
  \caption{The measured frequency ratio $(\omega_B/\omega_D)^2$ for a
  Bose gas with $7\times 10^4$ atoms after heating in the optical lattice.
  The rms axial width is given after 15 ms of time-of-flight. The inset shows the evolution of
  the axial width  vs. hold time in the optical lattice prior to excitation of the collective modes.}
  \label{fig4}
\end{figure}

To study the transition from the 1D quantum degenerate gas to the
1D thermal gas we have prepared atomic clouds in the optical
lattice with increasing non-condensed fraction but constant atom
number. We have loaded an initially pure Bose-Einstein condensate
of $7\times 10^4$ atoms into an optical lattice and trapped it
there for different hold times before exciting the collective
modes. During the hold period the condensate is subjected to
heating by off-resonant photon scattering and possibly technical
noise on the trapping fields. We find that together with the hold
time the axial width of the atomic cloud increases (see inset of
Fig. 4), whereas the radial size is unaffected. Since the
timescale for thermalization of the 1D gas is unknown it remains
unclear whether the cloud is in thermal equilibrium and we refrain
from calculating a temperature from the rms axial width of the
cloud. Figure 4 shows the measured ratio $(\omega_B/\omega_D)^2$
as a function of the rms axial width of the cloud. For increasing
width the ratio $(\omega_B/\omega_D)^2$ approaches 4, which is the
value for a classical noninteracting gas. From a simple estimate
we deduce that the 1D gases are in a collisional regime along the
axial direction, since $N_{1D} \cdot (1-{\cal T} ) \gg 1$, where
${\cal T}$ is the transmission coefficient for a 1D collision of
two bosons \cite{Olshanii1998} and $N_{1D} \sim  70$ is the number
of atoms in a 1D tube.

In conclusion, we have realized both thermal and quantum
degenerate gases in one dimension and investigated their physics
by measuring the low-lying collective excitations. Our
measurements show that the properties of the 1D ground state are
extremely sensitive to thermal excitations and finite temperature
effects must be taken into account when studying 1D gases, in
particular for the identification of the Tonks-Girardeau regime of
impenetrable bosons.

We would like to thank M. Greiner, J.P. Stucki, the workshops of
D-PHYS and of PSI, Villingen, for help during construction of the
experiment, C. Schori, G. Blatter, H.P. B{\"u}chler, and W.
Zwerger for insightful discussions, and SNF and SEP Information
Sciences for funding.


\begin{thebibliography}{}
\bibitem{remark1} The switch-off time for the optical lattice is
5\,$\mu$s and for the magnetic trap 300\,$\mu$s.
\end{thebibliography}
\end{document}